\documentclass[]{spie}  

\usepackage{amsmath,amsfonts,amssymb}
\usepackage{aas_macros}
\usepackage{subfig}
\usepackage{graphicx}
\usepackage[colorlinks=true, allcolors=blue]{hyperref}

\title{In-orbit anomaly and recovery of IXPE detector 2}

\author[a]{Fabio Muleri}
\author[b,c]{Luca Baldini}
\author[d]{Sudip Chakraborty}
\author[d]{Chen Chien-Ting}
\author[a]{Enrico Costa}
\author[a]{Ettore del Monte}
\author[a]{Alessandro Di Marco}
\author[e]{Kurtis L. Dietz}
\author[a]{Sergio Fabiani}
\author[a]{Riccardo Ferrazzoli}
\author[e]{Philip Kaaret}
\author[f]{Alberto Manfreda}
\author[e]{Stephen L. O'dell}
\author[g]{John Rankin}
\author[e]{Brian D. Ramsey}
\author[c]{Carmelo Sgrò}
\author[a]{Paolo Soffitta}
\author[d]{Douglas A. Swartz}
\author[e]{Allyn F. Tennant}
\author[h]{Alessio Trois}
\author[e]{Martin C. Weisskopf}

\affil[a]{INAF -- IAPS, via Fosso del Cavaliere, 100, Rome, Italy I-00133}
\affil[b]{Università di Pisa, Dipartimento di Fisica Enrico Fermi, Largo B. Pontecorvo 3, I-56127 Pisa, Italy}
\affil[c]{INFN-Pisa, Largo B. Pontecorvo 3, I-56127 Pisa, Italy}
\affil[d]{Science and Technology Institute, Universities Space Research Association, Huntsville, AL 35805, USA}
\affil[e]{NASA Marshall Space Flight Center, Huntsville, AL 35812, USA}
\affil[f]{INFN-Napoli, via Cintia, I-80126, Napoli, Italy}
\affil[g]{Dipartimento di fisica e scienze della terra, Università di Ferrara, Via Giuseppe Saragat, 1, Ferrara, Italy I-44122}
\affil[h]{INAF-OAC, Via della Scienza 5, I-09047 Selargius (CA), Italy}

\authorinfo{Further author information: (Send correspondence to F.M.)\\F.M.: E-mail: fabio.muleri@inaf.it, Telephone: +39 06 4993 4565}

\begin{document} 
\maketitle

\begin{abstract}
The Imaging X-ray Polarimetry Explorer (IXPE) is a joint NASA--ASI mission launched in December 2021 to measure the linear polarization of celestial X-ray sources in the 2--8~keV energy range. The payload comprises three identical telescopes, each combining grazing-incidence X-ray optics with a focal-plane photoelectric polarimeter based on the Gas Pixel Detector technology. On 14 April 2025, the second detector experienced an anomaly that significantly altered the electrical behavior of the ASIC responsible for reading out the photoelectron track images. In this contribution, we describe the effect of the anomaly on the instrumental response and the actions taken to mitigate its effects and recover the scientific observations of the instrument.
\end{abstract}


\section{INTRODUCTION} \label{sec:intro}  

\begin{figure}
\centering
\includegraphics[width=0.9\linewidth]{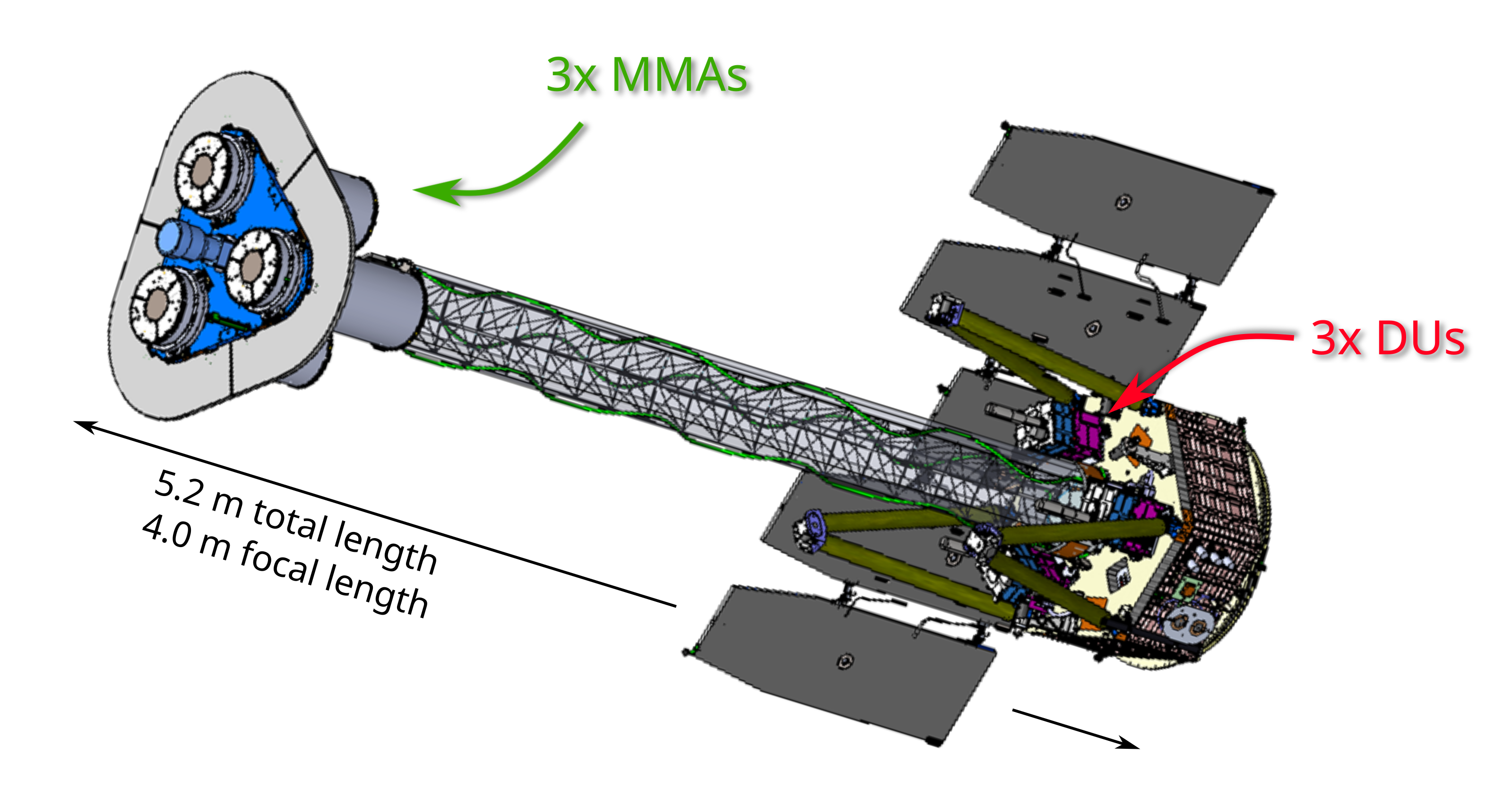}
\caption{Sketch of IXPE in orbit. The Mirror Module Assemblies (MMAs) are mounted on an extensible boom that is deployed in orbit to achieve a 4~m focal length. The focal plane detectors are mounted at the top of the spacecraft.}
\label{fig:ixpe}
\end{figure}

The Imaging X-ray Polarimetry Explorer (IXPE) is a small mission jointly developed by NASA and ASI\cite{Weisskopf2022}. The mission carries a payload comprising three identical, co-aligned telescopes dedicated to measuring the linear polarization of celestial sources in the soft X-ray band, over the 2--8~keV energy range (see Figure~\ref{fig:ixpe}). The mission is led by NASA's Marshall Space Flight Center, which also provided the ultra-light grazing-incidence mirror assemblies, achieving a point spread function better than 30~arcsec\cite{Ramsey2019,Ramsey2025}, with a contribution from Nagoya University for thermal filters. The focal plane instrumentation was provided by ASI, INAF, and INFN\cite{Soffitta2021}. The detectors are imaging photoelectric polarimeters based on the Gas Pixel Detector (GPD), developed in Italy since 2001\cite{Costa2001,Bellazzini2006,Bellazzini2007}. The GPD operates by imaging the track of the photoelectron emitted following the photoabsorption of an X-ray photon. The former is identified by the ion-electron pairs produced through ionization: the latter are drifted toward a Gas Electron Multiplier (GEM) and multiplied before begin collected on the top of a pixillated ASIC developed specifically for X-ray polarimetry. The measured track is analyzed to determine both the absorption point and the initial emission direction of the photoelectron, the latter being correlated with the polarization of the incident radiation. The energy of the absorbed photon is proportional to the total collected charge, while the trigger time provides its arrival time with a precision of a few microseconds. Each GPD is housed in a Detector Unit (DU)\cite{Baldini2021}, which also includes four calibration sources, providing both polarized and unpolarized radiation, as well as a filter for observations of very bright targets\cite{Muleri2018, Ferrazzoli2020}. Compared with its few predecessors, IXPE improves the sensitivity to polarization by approximately two orders of magnitude in the observing time required to detect a given polarization signal, while also enabling spatially, spectrally, and temporally resolved polarimetric measurements.

IXPE was launched on 9 December 2021 into a low-equatorial orbit at an altitude of approximately 600~km. Since the start of science operations on 11 January 2022, IXPE has detected X-ray polarization from tens of celestial sources spanning nearly all major classes of X-ray bright astrophysical objects. These observations have demonstrated that X-ray polarization is detectable in approximately two-thirds of the observed sources and provides powerful new observables for discriminating among competing theoretical models\cite{Liodakis2022,Doroshenko2022,Taverna2022,Krawczynski2022,DiGesu2023,Veledina2024}.

The scientific payload aboard IXPE has operated smoothly since launch. However, on 14 April 2025 an anomaly was detected in the second focal plane detector, designated det2 or DU2. This paper describes the anomaly, its impact on the detector performance, and the mitigation actions implemented to minimize its effects.

\section{THE ANOMALY OF det2}

\begin{figure}
\centering
\subfloat[\label{fig:vref}]{\includegraphics[width=0.9\linewidth]{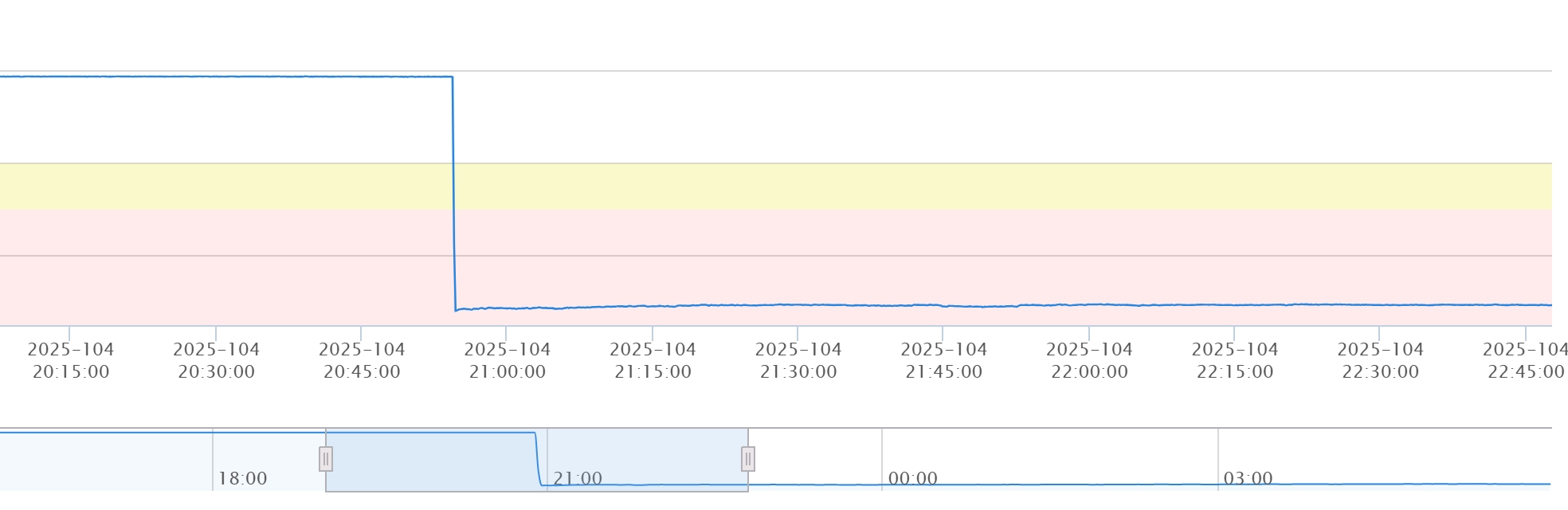}} \
\subfloat[\label{fig:counts}]{\includegraphics[width=0.9\linewidth]{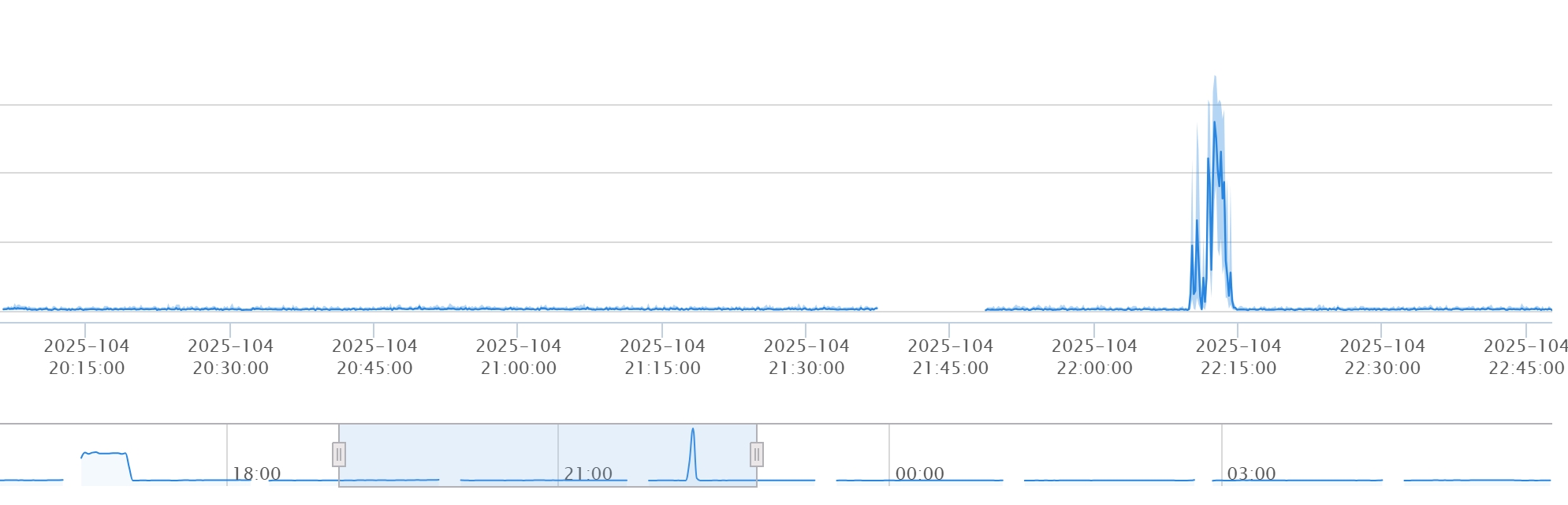}}
\caption{(a) Sudden drop of the output voltage $V_\text{ref}$ of the det2 GPD ASIC from 732~mV to 532~mV, recorded on 14 April 2025 at 20:54:01~UTC. (b) Trigger rate over the same time interval. The detector continued to operate after the anomaly, with an anomalous burst of triggers caused by noisy pixels occurring about 1 hour later.}
\label{fig:drop}
\end{figure}

On 14 April 2025 at 20:54:01~UTC, during a routine observation of the Kes~75 supernova remnant (observation ID 04002301), a sudden drop in one of the output voltages of the det2 GPD ASIC, denoted as $V_\text{ref}$, was recorded during the regular monitoring of the housekeeping parameters, sampled once per second. The voltage dropped from 732~mV to 532~mV (see Figure~\ref{fig:vref}). The GPD, which is based on a custom ASIC specifically developed for X-ray polarimetry\cite{Bellazzini2006}, continued to operate and trigger on incoming events after the anomaly. However, approximately 1~h~15~min after the anomaly, an anomalous burst of events triggered by noisy pixels was observed, lasting for about 5 minutes (see Figure~\ref{fig:counts}). The other two detectors aboard IXPE, which are completely independent, continued to operate normally and showed no anomalies.

\section{Consequences of the anomaly on det2 performance}

\begin{figure}
\centering
\includegraphics[width=1.0\linewidth]{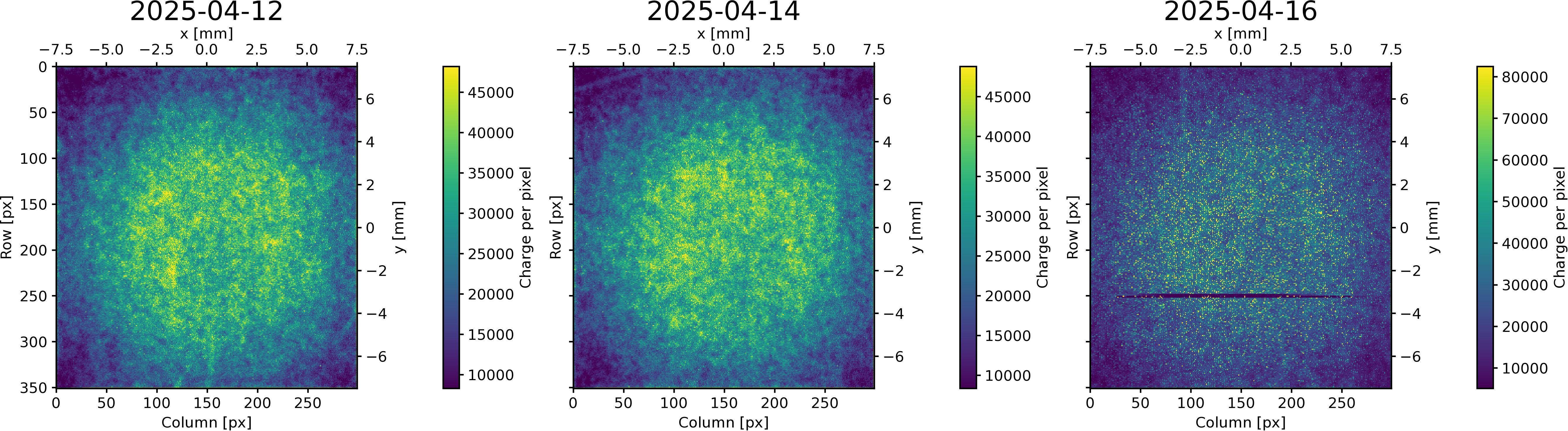}
\caption{Charge maps of det2 acquired on 12, 14, and 16 April 2025. The map acquired after the anomaly clearly shows that rows 248--251 became poorly responsive following the event on 14 April.}
\label{fig:chargemaps}
\end{figure}

The anomaly affected the detector performance in several ways. First, rows 248--251 of the ASIC, which consists of a $300\times352$ array of hexagonal pixels with a 50~$\mu$m pitch, became poorly responsive. This is clearly visible in Figure~\ref{fig:chargemaps}, which compares the so-called charge maps acquired before and after the anomaly. These maps are obtained by summing, pixel by pixel, the charge collected from a large sample of photoelectron tracks while the entire detector is uniformly illuminated by an X-ray source. In this case, the illumination is provided by the onboard calibration source C (calC), which emits X-rays at approximately 5.9~keV through the decay of a $^{55}$Fe radionuclide.

The reduced response of rows 248--251 to the input charge prevents the complete collection of the photoelectron image and, consequently, an unbiased reconstruction of the photoelectron's initial emission direction. To mitigate the impact on the polarimetric response, events collected in the region surrounding the affected ASIC rows must be excluded from the scientific analysis, resulting in a corresponding reduction of the amount of usable data. In practice, this is achieved by applying an appropriate spatial filtering to remove events near the affected rows and when generating exposure maps for det2 observations. For point-like sources, the resulting loss is small because the affected rows lie near the edge of the region where the source is imaged. During observations, the IXPE pointing follows a Lissajous dither pattern, causing a point-like source to trace a nearly circular spot with a diameter of approximately 1.5~mm in the central region of the detector sensitive area. The exact position of this spot depends on both the spacecraft pointing and the relative alignment of the three IXPE telescopes: the outer edge of the spot typically lies away from the affected rows.

\begin{figure}
\centering
\includegraphics[width=0.8\linewidth]{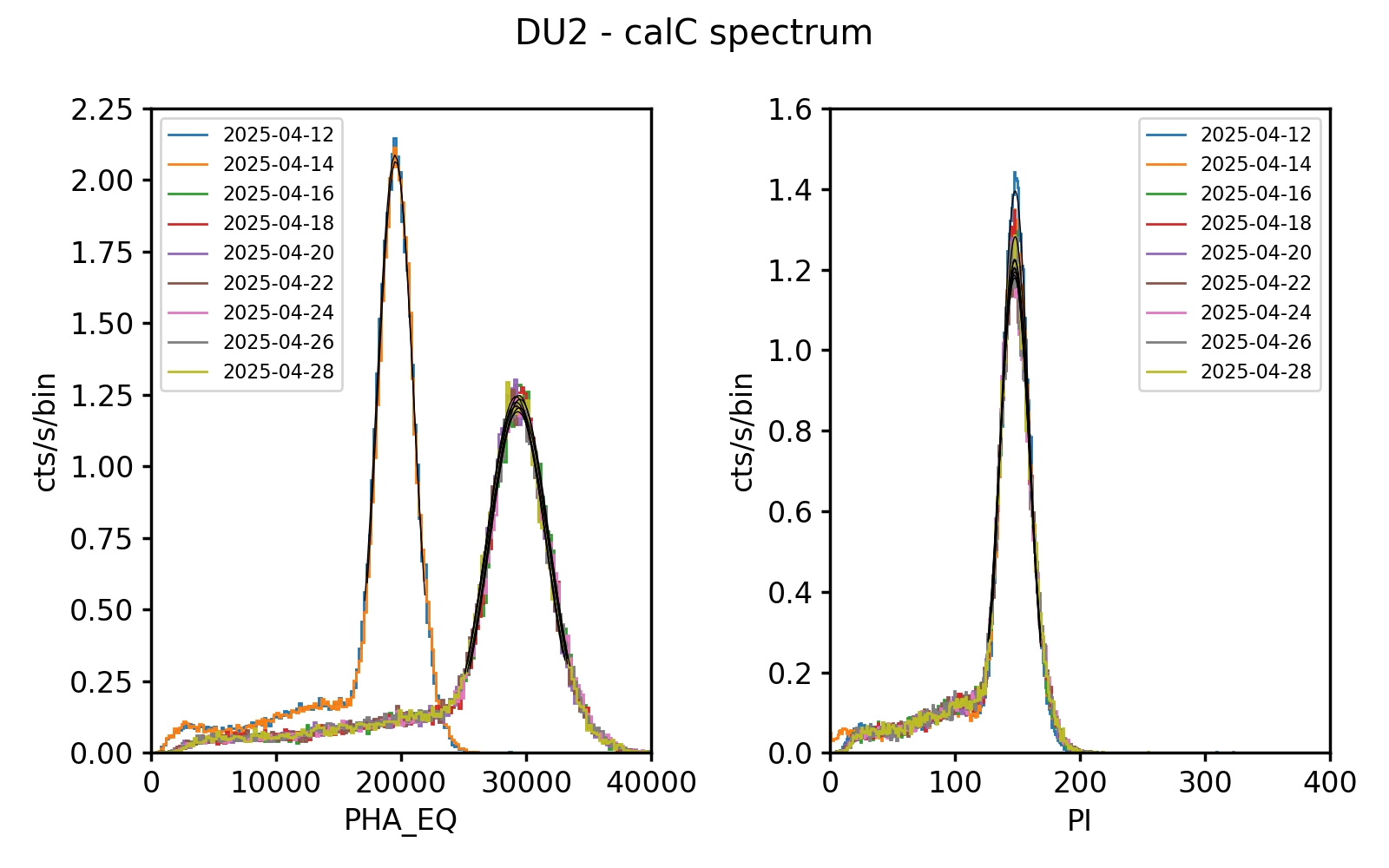}
\caption{Spectra of calibration source~C measured in uncalibrated (PHA\_EQ, left panel) and calibrated (PI, right panel) units at different times before and after the anomaly.}
\label{fig:calc_spectra}
\end{figure}

Another consequence of the anomaly was an increase in the \emph{average} electronic gain of \emph{all} the pixels in det2 (we will see in Section~\ref{sec:recal} that also the gain slope and offset of \emph{individual} pixels changed). This is illustrated in Figure~\ref{fig:calc_spectra}, which shows the spectra of calibration source~C (calC) measured before and after the anomaly in both uncalibrated (PHA\_EQ, left panel) and calibrated (PI, right panel) units. The latter differ from the former because they are corrected for the known sources of gain variability, including a renormalization based on the detector gain measured every two days using the onboard calibration sources\cite{DiMarco2022d, Rankin2023b}. The spectra in uncalibrated units clearly show the sudden increase in gain following the anomaly. Thanks to the onboard calibration sources, this increase can be measured accurately and is found to be $1.4931\pm0.0064$. As a consequence, the photoelectron image is rescaled by this factor before the track reconstruction. Nevertheless, the correct photon energy is fully recovered through the energy calibration applied at later stages of the science data processing pipeline, as demonstrated by the calibrated spectra shown in the right panel of Figure~\ref{fig:calc_spectra}. It is also worth noting that the energy resolution slightly degraded after the anomaly, changing from approximately $17.5\%$ to $18.5\%$ at 5.9~keV. However, it has remained stable since then and is still well within the IXPE requirement of 25\%.

\begin{figure}
\centering
\includegraphics[width=0.8\linewidth]{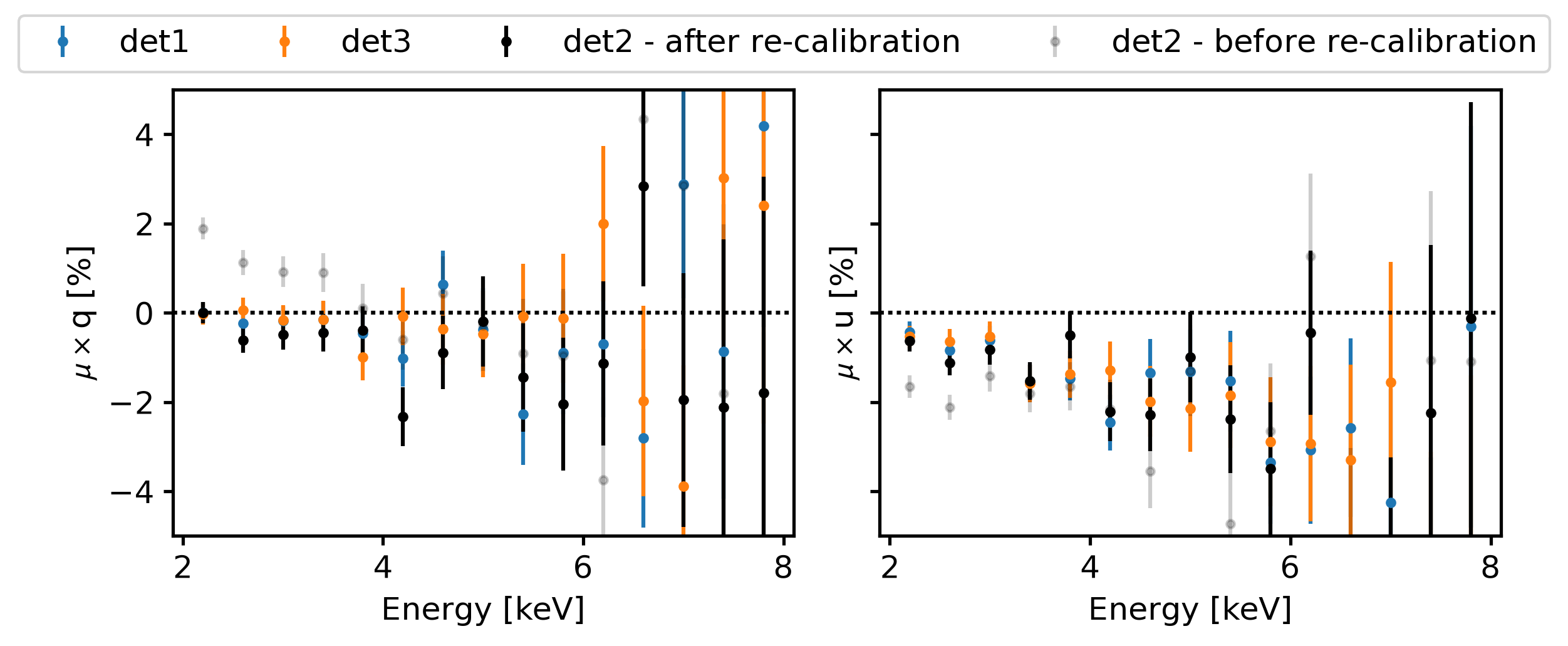}
\caption{Comparison of the polarimetric response of det1, det2, and det3, expressed in terms of the calibrated Stokes parameters $q$ and $u$\cite{Muleri2022}, as a function of energy. Stokes parameters are not normalized by the modulation factor. The post-anomaly values for det2 are shown in gray, while the values obtained after the in-flight recalibration (see Section~\ref{sec:recal}) are shown in black.}
\label{fig:stokes_v06}
\end{figure}

Finally, the anomaly affected the polarimetric response of det2 in two respects. The first effect was observed at low energies ($<4$~keV). Figure~\ref{fig:stokes_v06} compares the polarimetric response of det1, det2, and det3, expressed in terms of the calibrated Stokes parameters $q$ and $u$\cite{Muleri2022}, as a function of energy for an observation of the celestial source Cyg~X-2 (ObsID: 04250501, performed in May~2025). The post-anomaly values measured by det2 (gray points) deviate by several percent from those measured by det1 and det3 (blue and orange points, respectively). This translates into biases of several percent in the reconstructed polarization degree and, consequently, in the polarization angle. 

\begin{figure}
\centering
\includegraphics[width=0.95\linewidth]{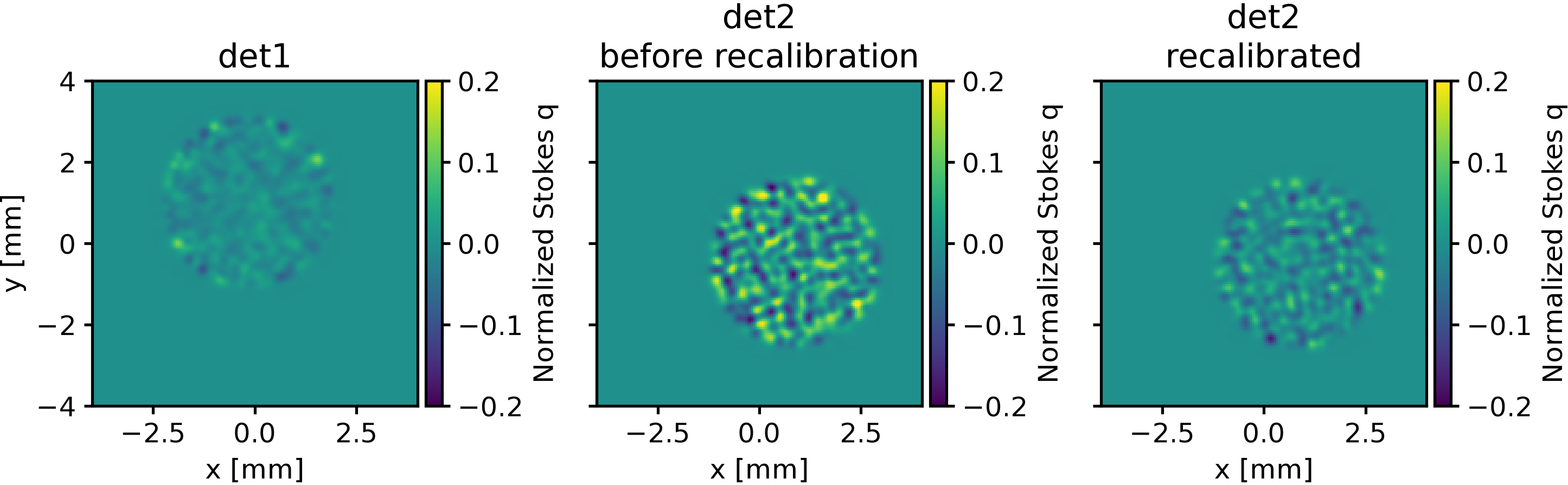}
\caption{Maps of the calibrated Stokes parameter $q$ over the detector plane for det1 (left panel) and det2 before (central panel) and after (right panel) the in-flight recalibration. The observation corresponds to Cyg~X-2 (ObsID: 04250501, May~2025), and the data are integrated over the 2--4~keV energy range. The offset between the det1 and det2 image locations is due to the images being presented in physical detector coordinates.}
\label{fig:stokes_maps_corr}
\end{figure}

\begin{figure}
\subfloat[]{\includegraphics[totalheight=7cm]{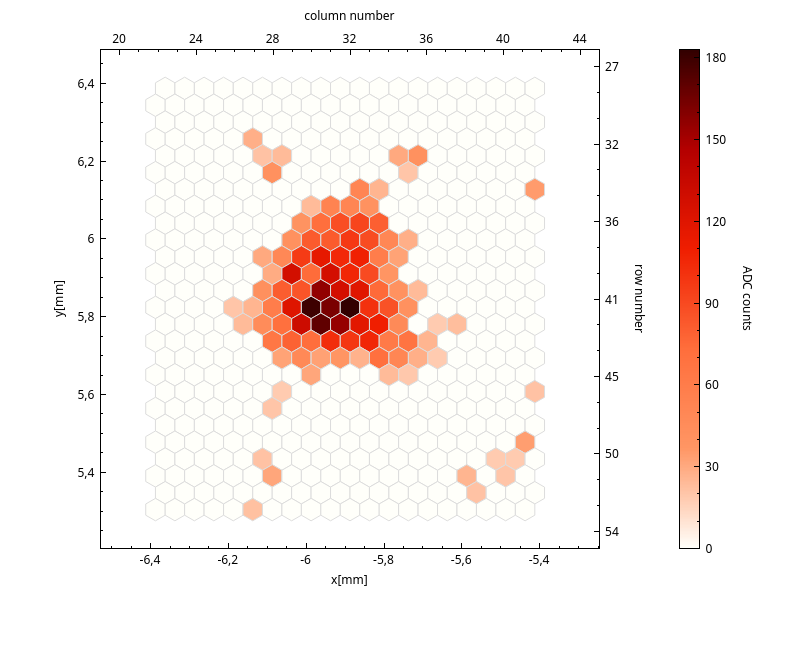}}
\hspace{3mm}
\subfloat[]{\includegraphics[totalheight=7cm]{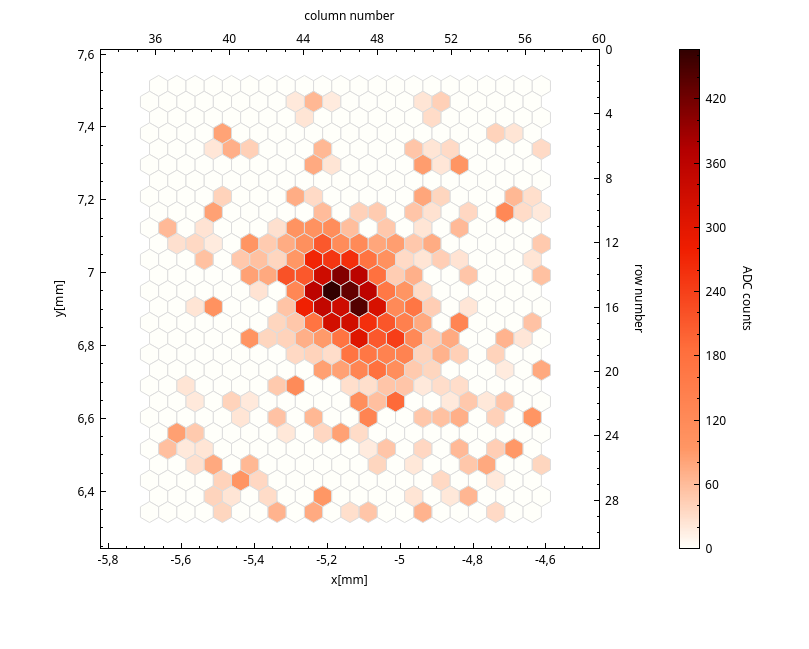}}
\caption{Typical photoelectron track recorded by det2 before (a) and after (b) the anomaly.}
\label{fig:tracks}
\end{figure}

Moreover, the effect depends on the photon absorption position within the detector. This behavior is illustrated in Figure~\ref{fig:stokes_maps_corr}, which shows, as an example, the detector-plane maps of the calibrated Stokes parameter $q$ measured by det1 and det2 during the same observation of Cyg~X-2. Although the source is point-like, it illuminates an extended region of the detector because of the spacecraft dithering. The maps are computed in the 2--4~keV energy range. The expected response is represented by det1 (left panel), where the calibrated Stokes parameter is spatially uniform, apart from the statistical fluctuations expected from the finite number of events. In contrast, the post-anomaly response of det2 (central panel) exhibits much larger spatial variations, exceeding $\pm10\%$. Both effects can ultimately be attributed to an increase in the electronic noise affecting the photoelectron images read out by the ASIC. Examples of typical photoelectron tracks recorded before and after the anomaly are shown in Figure~\ref{fig:tracks}.

The impact of the det2 anomaly on the polarimetric response is significant, as the introduced bias is much larger than the typical statistical uncertainty of IXPE observations. For this reason, an in-flight recalibration of the det2 polarimetric response was mandatory. The procedure developed for this purpose is described in the following section.

\section{RECALIBRATION OF THE POLARIMETRIC RESPONSE} \label{sec:recal}

The calibration of the IXPE detectors corrects the polarimetric response in two main steps. The first is applied before track reconstruction, when the charge collected by each pixel in the photoelectron image is corrected to equalize the response of the pixel to the average pixel response\cite{Rankin2023}. The pixel-by-pixel correction parameters are derived from (long) measurements with calibration sources, which allow the response of each pixel to different input charge values to be measured. The response of each pixel is then fitted with a linear function to determine its gain slope and offset\cite{Rankin2023}.

The pixel equalization parameters for det2 were remeasured after the anomaly using observations of the onboard calibration source ``calD'', which uniformly illuminates the detector with 1.7~keV photons produced by fluorescence from a silicon target\cite{Muleri2018, Ferrazzoli2020}. CalD is observed every two days as part of the routine detector gain calibration. To achieve sufficient statistics for every pixel of the GPD ASIC, we stacked all observations acquired over several months, from 15 July 2025 to 15 November 2025. 

\begin{figure}
\centering
\includegraphics[width=\linewidth]{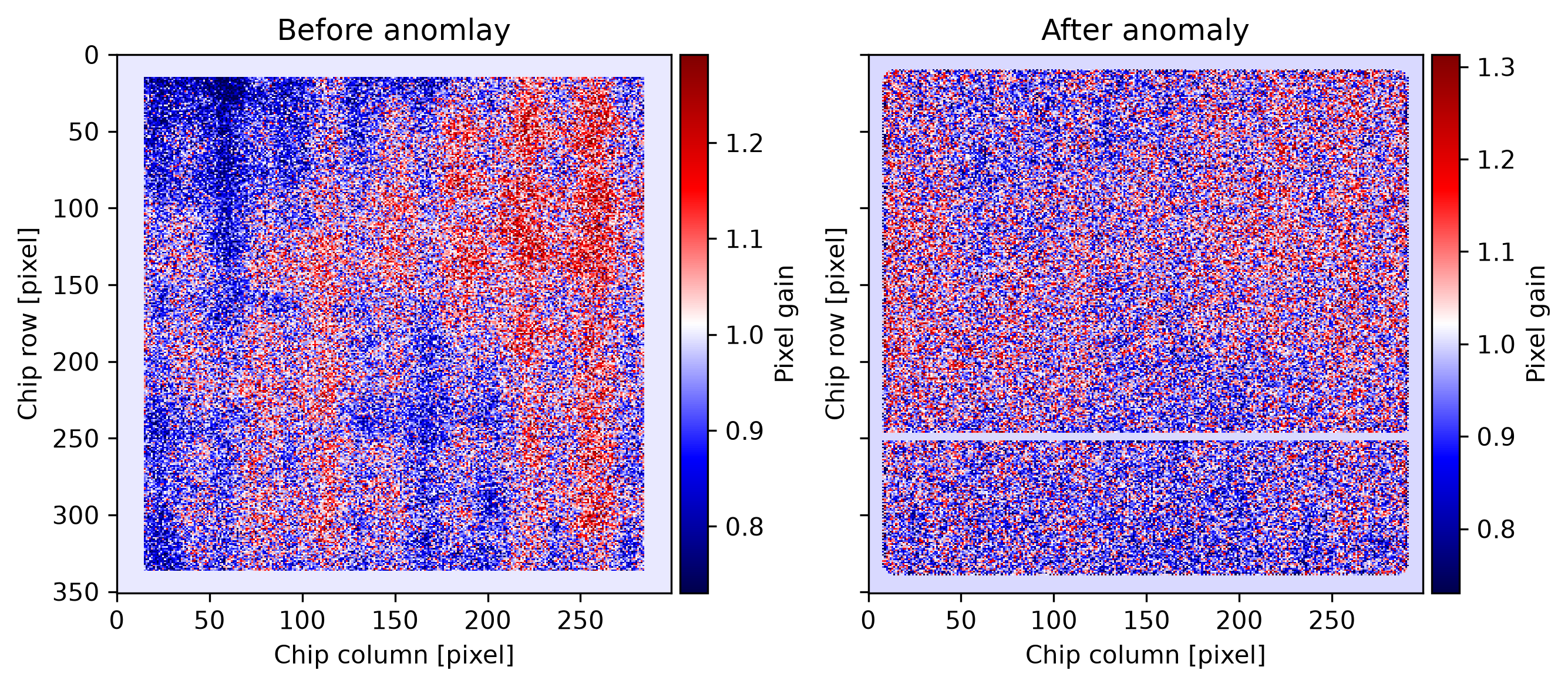}
\caption{Comparison of the pixel gain-slope maps measured before the anomaly (left) and after the anomaly (right).}
\label{fig:pixeq_maps}
\end{figure}

\begin{figure}
\centering
\includegraphics[width=\linewidth]{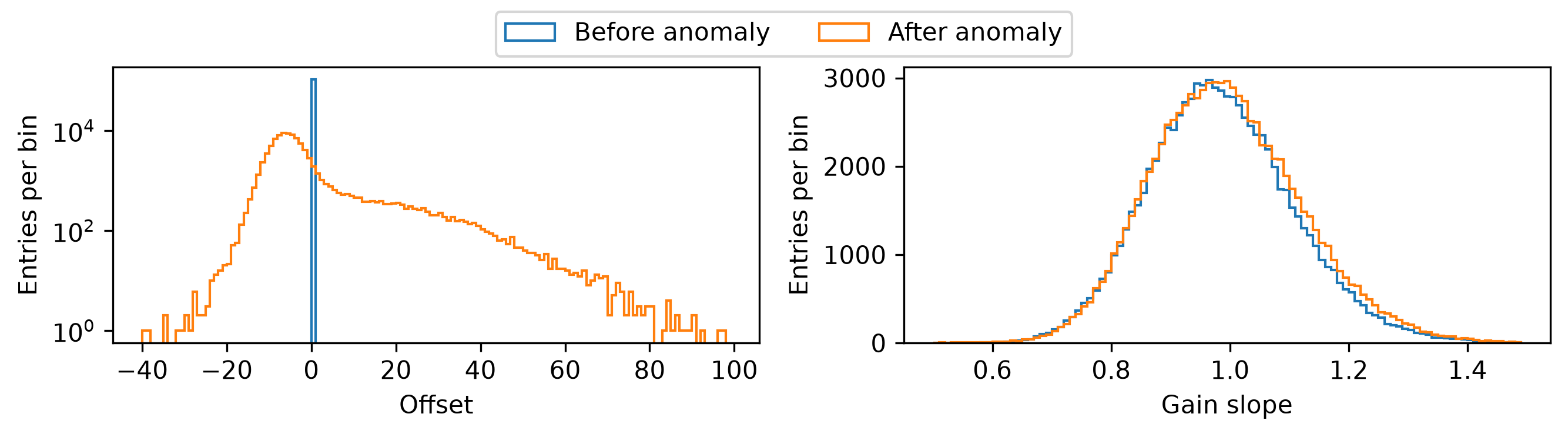}
\caption{Distributions of the pixel gain offset (left) and gain slope (right) before and after the anomaly. The most evident difference is the appearance of a significant gain offset, which was consistent with zero before the anomaly.}
\label{fig:pixeq_hist}
\end{figure}

The post-anomaly gain-slope map is compared with the pre-anomaly map, obtained during the ground calibration campaign, in Figure~\ref{fig:pixeq_maps}. The large-scale structures are similar in the two maps: the vertical stripes are caused by gain non-uniformities of the GEM, which were unaffected by the anomaly. However, the pixel-to-pixel variations are noticeably enhanced after the anomaly, although the overall range of gain-slope values remains comparable (see also Figure~\ref{fig:pixeq_hist}). This is not the case for the gain offset. Before the anomaly, the offset was consistent with zero, as the GPD ASIC performs real-time pedestal subtraction. After the anomaly, however, a significant non-zero offset is measured for a substantial fraction of the pixels. The distributions of the gain offset and gain slope before and after the anomaly are compared in Figure~\ref{fig:pixeq_hist}.

The second stage of the IXPE calibration to polarimetric response is applied after track reconstruction. The reconstructed emission direction of the photoelectron is affected by an energy- and position-dependent systematic bias, which produces a signal that mimics polarization even for unpolarized radiation. This bias, referred to as \emph{spurious modulation}, was characterized during the ground calibration campaign and is subtracted from observations of celestial sources, allowing the scientific analysis to be performed on calibrated quantities that closely reproduce the response of an ideal detector\cite{Rankin2022}. Following the det2 anomaly, this bias had to be remeasured. Unfortunately, the onboard calibration sources do not provide a sufficiently fine energy sampling to characterize the energy dependence of the spurious modulation, as the available unpolarized calibration lines are only at 1.7 and $\sim$5.9~keV.

We therefore adopted a different approach based on observations of celestial sources. The source polarization, which is generally unknown and may be intrinsically variable, is measured independently with det1 and det3, which were unaffected by the anomaly. The intrinsic source polarization signal is then removed from the uncalibrated response of det2, so that the residual signal, by definition, corresponds to the spurious modulation. Since the energy spectrum of a celestial source is essentially continuous, the dependence of the spurious modulation on energy can be measured by binning the events in energy. Achieving a sufficiently small statistical uncertainty requires a large number of events. For this reason, the moderately bright X-ray source Cyg~X-2 was observed for approximately two months, from December~2025 to January~2026, using raster pointings to cover the central region of the detector field of view.

\begin{figure}
\centering
\includegraphics[width=0.9\linewidth]{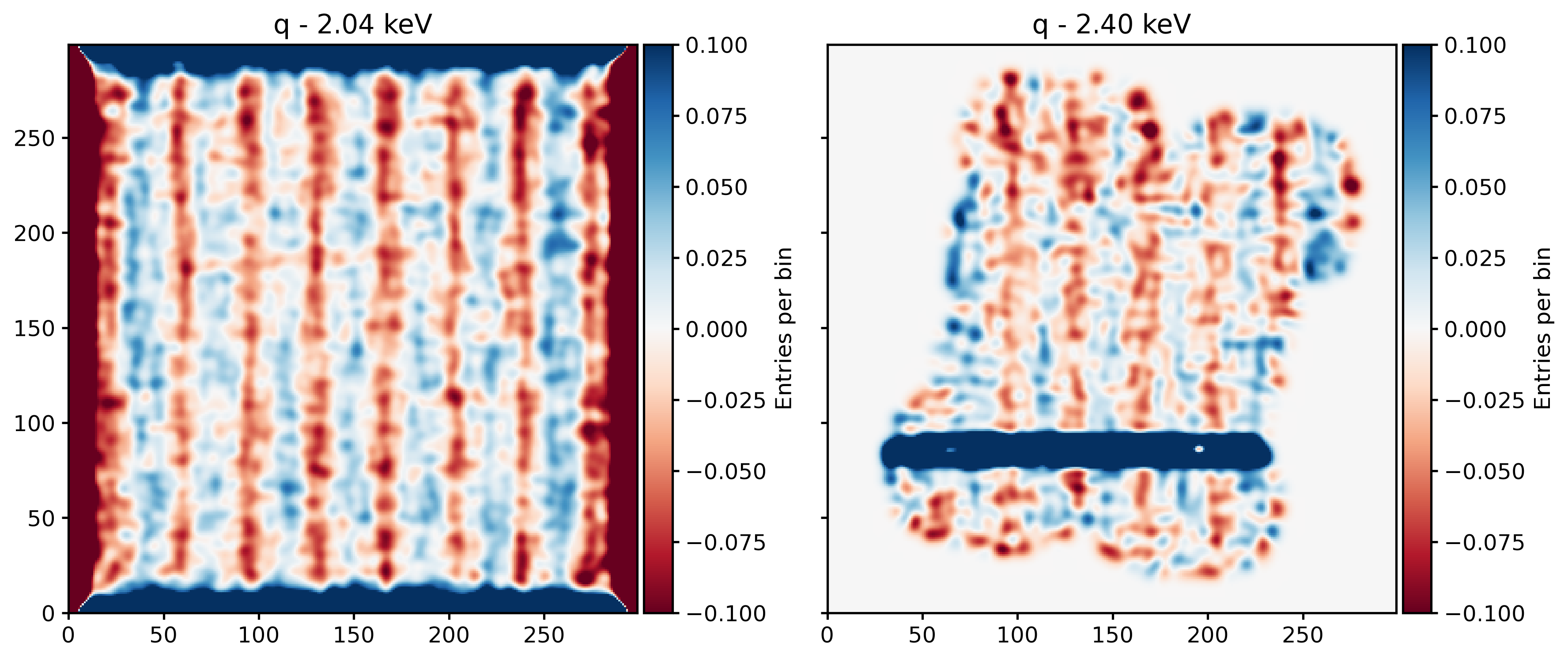}
\caption{Comparison of the Stokes $q$ spurious-modulation maps at the lowest available energy before the anomaly (left, ground calibration) and after the anomaly (right, in-flight calibration). Both maps are smoothed with a Gaussian kernel having a standard deviation of $3\times3$ pixels.}
\label{fig:maps_q}
\end{figure}

An example of the spurious modulation map obtained with this procedure is shown in Figure~\ref{fig:maps_q}. The figure compares the Stokes $q$ map measured during the ground calibration campaign with that obtained from the Cyg~X-2 observations in the lowest-energy bin, where the spurious modulation is largest (line emission at 2.04 keV for the ground calibration and continuum emission over the 2-3 keV band with an average energy of 2.40 keV for the post-anomaly calibration). Both maps have been smoothed with a Gaussian kernel to reduce statistical fluctuations. The comparison highlights that the large-scale pattern associated with the GEM\cite{Baldini2021} is preserved after the anomaly, while significant local variations in the spurious modulation are clearly visible.

The effectiveness of the in-flight recalibration was assessed by applying the new calibration products, namely the updated pixel equalization parameters and the in-flight spurious modulation maps, to an independent observation that was not used during the calibration process. The resulting polarimetric response as a function of energy is shown by the black points in Figure~\ref{fig:stokes_v06}, together with the gray points obtained using the original processing based on the ground calibration products. The systematic bias of det2 with respect to det1 and det3 is substantially reduced, and the det2 measurements become statistically consistent with those of the other two detectors. Likewise, the residual spatial variations of the normalized Stokes parameters are significantly reduced, as illustrated in Figure~\ref{fig:stokes_maps_corr}. The recalibrated map shown in the right panel exhibits much smaller spatial variations than those present before the recalibration (middle panel).

\section{CONCLUSIONS}

An electronic anomaly occurred on 14 April 2025 in the second detector of IXPE. The anomaly affected the detector performance in several respects, including the loss of a small portion of the field of view, an increase in the electronic gain across the entire detector, and a slight degradation of the energy resolution. The most significant impact, however, was on the polarimetric response, which is particularly sensitive to even small variations in the uniformity of the pixel response.

Fortunately, the combination of onboard calibration data and dedicated observations of bright celestial sources made it possible to perform an in-flight recalibration of det2, although with some unavoidable trade-offs. On the one hand, practical limitations in both the brightness of celestial calibration targets and the observing time available for calibration prevented the production of spurious modulation maps with the same statistical quality and spatial coverage as those obtained during the ground calibration campaign. Also, the increased electronic noise in the photoelectron images degrades the efficiency of the background-rejection algorithms\cite{DiMarco2023}. This effect has been significantly mitigated applying a new set of parameters for the background rejections (as reported in the appendix of Di~Marco\cite{DiMarco2026}) and included in the IXPE caldb and in the version 3.2 of the dedicated software\footnote{https://github.com/aledimarco/IXPE-background}. The post-anomaly calibration products can be applied to all observations acquired after the anomaly, which have consequently been reprocessed and publicly released for scientific analysis.

The exact origin of the anomaly remains difficult to determine. The science data collected at the time of the event indicate the presence of a saturated energy deposit, possibly associated with the passage of a highly ionizing particle along an unfavorable trajectory through the detector. Regardless of its origin, the event appears to be exceedingly rare: this is the first occurrence after nearly four years of operation of the three IXPE detectors in orbit, and no similar event was ever observed during the more than twenty years of ground development and testing of the GPD.

\section*{Acknowledgements}

The Imaging X-ray Polarimetry Explorer (IXPE) is a joint US and Italian mission.  The US contribution is supported by the National Aeronautics and Space Administration (NASA) and led and managed by its Marshall Space Flight Center (MSFC), with industry partner Ball Aerospace (now, BAE Systems).  The Italian contribution is supported by the Italian Space Agency (Agenzia Spaziale Italiana, ASI) through contract ASI-OHBI-2022-13-I.0, agreements ASI-INAF-2022-19-HH.0 and ASI-INFN-2017.13-H0, and its Space Science Data Center (SSDC) with agreements ASI-INAF-2022-14-HH.0 and ASI-INFN 2021-43-HH.0, and by the Istituto Nazionale di Astrofisica (INAF) and the Istituto Nazionale di Fisica Nucleare (INFN) in Italy.  This research used data products provided by the IXPE Team (MSFC, SSDC, INAF, and INFN) and distributed with additional software tools by the High-Energy Astrophysics Science Archive Research Center (HEASARC), at NASA Goddard Space Flight Center (GSFC).

\bibliography{References.bib}
\bibliographystyle{spiebib} 

\end{document}